# Working with Bill Kruskal: From 1950 Onward

Leo A. Goodman

## 1. INTRODUCTION

Bill Kruskal and I arrived at the University of Chicago at about the same time, a very long time ago, in time for the beginning of the 1950–1951 academic year. We became colleagues and very good friends, and we worked together very harmoniously and productively as colleagues, and also as co-authors, over a very long period of time. We started to work together in the early 1950s on the introduction and development of various measures of association for the analysis of cross-classified categorical data, and we published our first joint article on this subject in 1954, followed by a series of three other joint articles on the subject in 1959, 1963 and 1972; and the four articles were then brought together in a single volume in 1979. Bill and I worked on the first article—the core article—on and off for about two years before we submitted it for publication, and the series of four articles evolved over a 20-year period. The 1979 volume appeared in print 25 years after the publication of the first article; and now more than 50 years have gone by since the first article was published. Yes, a very long time has gone by.

I shall describe here some of the experiences that Bill and I shared over the years, from the early 1950s until 1987, when I retired from the University of Chicago (UChicago) to take up work at the University of California at Berkeley (UCBerkeley), and I shall also comment briefly here on some experiences shared from 1987 onward. The experiences described here will make clear some of Bill's very special—wonderfully special—characteristics. He was a wonderful person.

## 2. MEASURES OF ASSOCIATION

In a conversation that Bill had with Sandy Zabell, which was published in the 1994 *Statistical Science*, Bill said that the joint work that he and I had done on measures of association for cross-classifications grew out of a conversation that we had at a New Year's Eve party that Bill and I happened to attend at The Quadrangle (Faculty) Club. Our conversation at the party was about our earlier experiences serving as statistical consultants after we arrived at the university. As beginning faculty members, Bill had been asked to serve as a statistical consultant to Bernard Berelson in the Graduate Library School, and I had been asked to serve as a statistical consultant to Louis Thurstone in the Psychology Department.

Berelson was the dean of the Graduate Library School at that time and later became the president of the Population Council. He also was an important figure in the social and behavioral sciences at that time, and later became an even more important figure. Thurstone was a distinguished professor in the Psychology Department where he was the founder and director of the Psychometric Laboratory. He had been instrumental in the development of the field of psychometrics, and was at that time the major figure in the development of factor analysis. (By the way, as a very young, beginning assistant professor, I thought it passing strange that I had been asked to serve as a statistical consultant to the great L. L. Thurstone.)

Well, the conversation that Bill and I had at that party took place some time after Bill had met with Berelson and some time after I had met with Thurstone and some other members of his Psychometric Laboratory. Bill and I were describing to each other what happened when he met with Berelson and I met with the Thurstone group, and we observed in this conversation that the kinds of statistical problems with which Berelson was concerned and the kinds of statistical problems with which the Thurstone group was concerned could be viewed as problems concerning the measurement of association for cross classifications. We discovered that each of us had been independently thinking about similar kinds of questions. So, right then and there, at that party, Bill and I joined forces, and we were off and running. Incidentally, I would guess that it was Bill who had engaged me (rather than I who had engaged him) in this conversation about our work. I doubt that, as a young bachelor at that time, I would have engaged anyone in a conversation about work at a party, especially at a New Year's Eve party.

*Leo A. Goodman is the Class of 1938 Professor, Department of Statistics and Department of Sociology, University of California, Berkeley, California 94720-1980, USA e-mail: lgoodman@berkeley.edu.*

After we completed work on our first joint paper and submitted it for possible publication in the *Journal of the American Statistical Association* (*JASA*), we had to wait, of course, for referees to write their reports and for the reports to reach us. The reports finally arrived, and each of the reports stated, among other things, that the manuscript should be shortened: the main referee's report stated that the manuscript should be reduced by 50 percent! Right after reading these reports, I told Bill that I thought that the main ideas and results in our manuscript could be presented in a revised, shorter version, but these ideas and results would be much harder for the *JASA* reader to grasp in the shorter version and it would ruin the paper with respect to its accessibility. (All the efforts that Bill and I had made to make our work accessible to a very wide audience would come to naught in the shorter version.) I also told Bill that to satisfy the referees and the *JASA* editor, I would reluctantly be willing to shorten the manuscript by 50 percent. Fortunately, Bill did not accept my proposal, and he then wrote a very long, detailed letter to the editor explaining why the manuscript should not be shortened at all—why it should be published as is. The editor, after reading Bill's letter, accepted the manuscript for publication as is. (By the way, just in case the reader might be curious about this, I note here that the editor of *JASA* at that time was Allen Wallis, who was also at that time the first chairman of our nascent Department of Statistics. Also, just in case the reader might be interested in this too, the length of the article was 33 printed pages in *JASA*; the lengths of the second, third and fourth articles in the series were 41, 55 and 7 printed pages, respectively, in *JASA*.)

In 1979, the Institute for Scientific Information (ISI) informed Bill and me that our first article had been selected as a Citation Classic, and we were invited to write a commentary on that article, which the ISI published in *Current Contents, Social and Behavioral Sciences*. It turns out that, according to the ISI, there are about 1,060 citations of that article from the time of its publication in 1954 until now. In recent years, the average number of citations of the article per year is now even more than in earlier years. (The many citations of this one article appear in a seemingly boundless range of different articles in journals that cover a seemingly boundless range of different fields of study.) Also, it turns out that, according to the ISI, there are about 1,800 citations in total of the four articles that Bill and I wrote on measures of association and of the 1979 volume in which the four articles were brought together.

In the foreword to the 1979 volume (Goodman and Kruskal, 1979), Steve Fienberg commented as follows on our exposition in the core article:

> Because of their clarity of exposition, and their thoughtful statistical approach to such a complex problem, the guidance in this paper is as useful and important today as it was on its publication 25 years ago.

Now, more than 50 years have gone by since our first joint article was published, and we might ask again about the usefulness and importance of this article at the present time. Well, I haven't carried out a study to try to answer this question, but I did pick up a newly published textbook (copyright 2006) on statistics for the social sciences, and in the textbook's chapter on "Measuring Association in Contingency Tables" there were sections on Goodman and Kruskal's gamma and on Goodman and Kruskal's lambda, and also references to the Goodman–Kruskal tau and the Goodman–Kruskal uncertainty coefficient. These measures of association are also still being used in some of the major statistical computer packages.

We also find, of course, that obliteration by incorporation often takes place, and so, for example, the Goodman–Kruskal gamma will now often be referred to simply as "gamma," and similarly for the Goodman–Kruskal lambda and the Goodman–Kruskal tau. Incidentally, the so-called Goodman–Kruskal uncertainty coefficient, which was referred to in the newly published textbook, was not one of the measures of association considered in our series of articles. [It seems to me that attaching the Goodman–Kruskal name to this particular measure might be viewed as an example of what I would call "incorporation by association." In addition, this attachment of the Goodman–Kruskal name to this particular measure can also serve as a good example of "Stigler's Law of Eponymy" (see Stigler, 1999).]

I am absolutely certain that if Bill had accepted the proposal that I had *reluctantly* made right after reading those referees' reports, and if he had not written that very long, detailed letter to the *JASA* editor, the revised shortened paper that we would have written and had published in *JASA* would definitely not have had an effect in any way comparable to the effect that our 1954 *JASA* article actually has had.

In the 1994 *Statistical Science* conversation that Bill had with Sandy Zabell, Sandy asked Bill how did he and I interact when we were working on our measures of association series of articles. Bill responded as follows:

> Oh, we exchanged draft statements, we talked on the phone and in person. We got after this epistemological issue of interpretability. It was well hashed out between us. Then we got into relevant sampling theory and tried to write it up in an accessible way. I remember that while we were doing this, Leo spent a year in England at Cambridge...[and] we [Bill and his wife Norma] visited Leo [there], moving from one draft to the next there in England. That was great.

I would have added at least one exclamation mark at the end of Bill's last sentence above, and I think that Bill would also have done that, except that the use of exclamation marks didn't seem to have a place in Bill's writing style.

## 3. SOCIAL SCIENCES DEANSHIP

Bill was appointed Dean of the Social Sciences Division at the University of Chicago in time for the beginning of the 1974–1975 academic year. All previous deans of the Social Sciences Division had been selected from among the faculty members who were in the various departments of that division. Bill was *not* a faculty member in such a department. (The Department of Statistics was in the Division of the Physical Sciences, not in the Social Sciences Division.) Here is how I think it came about that Bill was selected even though he was not a member of the division:

Early in 1974, many faculty members in the Social Sciences Division were aware of the fact that a new dean needed to be selected. Various departments in the division had their own candidates whom they were promoting. I was aware of all this because I was a member of the Social Sciences Division (in the Sociology Department), as well as a member of the Physical Sciences Division (in the Statistics Department), and I had my own candidate whom I was promoting. When the faculty advisory committee was formed to advise the administration about the selection of the dean, I spoke to several members of the committee to promote the idea of selecting Bill to be invited to be the dean even though he was not a member of the division. The points I made in favor of selecting Bill had to do with his wide interests in topics related to the social sciences, his work and its relationship to the social sciences, his character, his personality, and so forth. I had the impression at that time that several members of the advisory committee and some other members of the Social Sciences Division liked this idea and that some of them then helped to promote it.

Bill was, as one might expect, an excellent dean—very thoughtful, very conscientious and very thorough. It was his view, at that time, that the most important part of the dean's job was to make recommendations about appointments, promotions and related matters. It sometimes seemed to him to be a challenge to come up with reasonable conclusions on the basis of the material provided by a department in support of a particular recommendation. In some cases where he thought that a really careful study of a person's research was necessary in order to be able to come up with a reasonable conclusion and where he thought that I might be able to assist him in this study, he called on me. Here again, as earlier when we worked on our measures of association project, we worked together to carry out the necessary study. Also, during the ten years of Bill's deanship, he and I would get together at times to consider problems that he faced in his role as the dean, problems of the kind that he felt I could assist him in solving.

## 4. FROM 1987 ONWARD

I left UChicago at the end of 1986 and began working at UCBerkeley at the beginning of 1987. Bill and I continued to keep in touch. I would send him drafts of papers on which I was working and he would send me back helpful comments. He would also send me reprints of interesting articles of his that were being published.

Then, in 1992, there was the very sudden, tragic death of Bill's wife, Norma. Not long afterward, Bill moved into Montgomery Place, a retirement community in the UChicago neighborhood. He was able to have an active and interesting life there for a number of years, but then some health problems, which predated Bill's move into the retirement community, began steadily to develop into more serious health problems. At that point, Bill and I took on, in a certain sense, another joint project. This project was altogether different from our earlier joint project on measures of association. Here is what it was: When Bill started to have the more serious health problems, he often felt very frustrated by the contact that he was having with the doctors who were treating him. It frequently turned out to be the case that, to try to gain a clearer understanding of some medical issue pertaining to his case, he would ask the doctor a question—sometimes a question of a statistical character—and then, after listening to the doctor's response to his question, Bill would feel that the doctor's response was inadequate or unclear or wrong. Bill remembered that I had had difficulties dealing with doctors when I had been diagnosed with cancer earlier, way back in 1976.

One of the difficulties that I had had was that one set of doctors advocated one way to deal with the cancer and another set of doctors advocated a very different way. Another one of the difficulties that I had had was that, when I studied the medical literature recommended to me by one of the sets of doctors, I found that the method of treatment recommended in the abstracts of those articles was consistent with the method of treatment advocated by that set of doctors, but it seemed to me that the detailed medical and statistical evidence presented in the articles themselves did not warrant the recommendation presented in the abstracts.

Bill remembered these difficulties, and other difficulties as well, that I had had in that earlier time period. We again joined forces, long distance this time. We dealt with, as best we could, whatever difficulties came up for Bill over time—whatever he wanted to go over with me. As a man of experience, where the outcome had turned out to be pretty good for me, I of course was hoping for a similarly pretty good outcome for Bill, but, alas . . . .

As I said at the beginning of this comment, Bill was a wonderful person. His influence will stay with me until the end.